\documentclass[twocolumn,showpacs,preprintnumbers,amsmath,amssymb]{revtex4} 
\usepackage{bm}
\input{psfig.sty} 
\begin{document} 
 
\title{Distance, dissimilarity index, and network community structure} 
 
\author{Haijun Zhou} 
 
\affiliation{Max-Planck-Institute of Colloids and Interfaces, 
D-14424, Potsdam, Germany}

\date{February 11, 2003} 
 
\begin{abstract} 
We address the question of finding the community structure of a  
complex network. In an earlier effort [H.\ Zhou, {\em Phys.\ Rev.\ E} (2003)], the 
concept of network random walking is introduced and a distance measure defined.  
Here we calculate, based on this distance measure, the dissimilarity index  
between  nearest-neighboring  
vertices of a network and 
design an algorithm to partition these vertices into communities that are  
hierarchically  organized.  Each community is characterized by  
an upper and a lower  dissimilarity threshold. 
The algorithm is applied to several artificial and real-world networks,  
and excellent results are obtained. In the case of artificially generated 
random modular networks, this method outperforms the algorithm based on the 
concept of edge betweenness centrality. For yeast's protein-protein 
interaction 
network, we are able to identify many clusters that have well defined biological 
functions.  
\end{abstract} 
 
 
\maketitle

\section{Introduction} 
\label{sec:introduction} 
 
A graph (network) of vertices (nodes) and edges  
is a useful tool in describing the 
interactions between different agents of a complex system. For example if we 
want to analyze  protein-protein  physical interactions  in yeast  
{\em Saccharomyces cerevisiae} \cite{uetz2000}, we would like to  
denote each 
protein as a distinct vertex of a graph, and setup an edge between 
two vertices if the corresponding proteins  
have direct physical interactions. Many such kinds of networks are 
constructed in sociological, biological, and technological fields, 
and they usually have very complicated connection patterns. What one needs is a  
method that is capable of classifying vertices of a 
complex network into different clusters (communities).  
If a network is appropriately decomposed  into a series of functional  
units, (a) the 
structure of the network can be better understood  
 and the relationship between its different components 
will be clear, (b) the  principal function of each cluster can  
be inferred from the functions of its members,  and (c) 
possible functions for members of a cluster can be suggested by 
comparing the functions of other members. Network clustering 
techniques are therefore  
very important in the emerging fields of bioinformatics 
and proteomics. 
 
A good clustering method needs to satisfy two conditions: First, the inherent structure 
of the network should be reserved; Second, it should provide a quantified 
resolution parameter to mark the significance of the clusters obtained at each level of 
the partitioning process.  
The global organization of a network should already be identified at low resolutions  
and more and more fine structures  emerge as the resolving power is increased.  
 
Many existing methods \cite{wasserman1994,ravasz2002} 
only take account of local information of each vertex, 
such as number of nearest-neighbors shared with other vertices, 
number of vertex-independent paths to other vertices, etc.. Recently, 
Girvan and Newman \cite{girvan2002} suggested an elegant global algorithm 
which extended the concept of vertex betweenness centrality of Freeman 
\cite{freeman1977} also to edges. Their algorithm works iteratively 
by removing the current edge(s) of the highest 
degree of betweenness centrality. When applying to an ensemble of 
random modular networks, this algorithm greatly outperforms  
some conventional methods \cite{girvan2002}. On the other hand, it 
does not provide a parameter to quantify the differences  
between communities. 
 
In reference \cite{zhou2003} a Brownian particle is $``$introduced" into a  
network to $``$measure" the distances between vertices. In the 
present work, we extend the basic idea of \cite{zhou2003} 
by defining, based on this distance matrix,  
a quantity called the dissimilarity index 
between nearest-neighboring vertices. The dissimilarity index  
signifies to what extent two nearest-neighboring vertices would 
like to be in the same community. 
A hierarchical algorithm is then worked out; it takes use of information on the 
dissimilarity indices and  decompose a network into a hierarchical sequence of clusters. 
Each of the communities is  
characterized by an upper and a lower dissimilarity threshold. 
 
The method, which could work on unweighted as well 
as weighted networks, is applied to several artificial and real networks, and 
very satisfying results are obtained. For the case of random modular 
networks, the present algorithm outperforms the  
method of Girvan and Newman \cite{girvan2002}. When applying the 
algorithm to the protein-protein interaction network of yeast, we are able to 
identify  many protein clusters which have well defined biological functions. 
 
In section \ref{sec:measures} we review the distance measure of 
reference \cite{zhou2003} and define a dissimilarity index 
for each pair of nearest-neighboring vertices. A dissimilarity-index-based hierarchical  
algorithm 
is outlined in section \ref{sec:algorithm}, and  applied to two kinds of artificially   
generated 
networks and four real-world networks in section \ref{sec:application}. 
We conclude our work in section \ref{sec:conclusion}.

\section{Distance measure and  dissimilarity index} 
\label{sec:measures} 
 
In the opinion of Flake, Lawrence, and Giles \cite{flake2000}, 
a community in a (sub)graph should satisfy the requirement that each vertex's total 
intra-community interaction be stronger than the total interaction with other 
vertices in the (sub)graph. This turns out to be 
a very strong constraint. In this work, we weaken this condition and 
require only that a vertex should have stronger total 
interaction with other vertices of its own community  than with vertices 
of any another community of the (sub)graph.  
 
We consider a connected network of $N$ vertices and $M$ edges. The network's 
connection pattern is specified by the generalized adjacency matrix 
$A$. We assume that the value of each non-zero element of matrix $A$ (say $A_{i j}$) denotes the interaction strength between vertex $i$ and $j$.  
The distance, $d_{i j}$,  from vertex $i$ to vertex $j$ is  
defined as the 
average number of steps needed for  
a Brownian particle on this network to move from vertex $i$ to 
vertex $j$ \cite{zhou2003}. At each vertex (say $k$) the Brownian 
particle will jump  in the next step to a nearest-neighboring vertex (say $l$) 
with probability  $P_{k l}=A_{k l}/\sum_{m=1}^N A_{k m}$.  
The distance matrix thus defined is asymmetric (in general 
$d_{i j}\neq d_{j i}$), and it is calculated by 
solving $N$ linear-algebraic equations \cite{zhou2003}. 
 
Taking any vertex $i$ as the origin of the network, then the set  
$\{d_{i 1},\cdots, d_{i,i-1}, d_{i,i+1},\cdots, d_{i N}\}$ measures  
how far all the 
other vertices are located from the origin. Therefore it is 
actually a perspective of the whole network with vertex $i$ being  
the viewpoint.  
Suppose vertex $i$ and $j$ are  nearest-neighbors ($A_{i j}>0$), the 
difference in their perspectives about the network can be 
quantitatively measured. 
We define the dissimilarity index, 
$\Lambda(i,j)$, by the following expression: 
\begin{equation} 
\label{dsi} 
\Lambda(i,j)= {\sqrt{\sum\limits_{k\neq i, j}^N [d_{i k}-d_{j k}]^2} 
\over (N-2)}. 
\end{equation} 

If two nearest-neighboring vertices $i$ and $j$ belong to the same 
community, then the average distance $d_{i k}$  
from $i$ to 
any another vertex $k$ ($k\neq i,j$) will be quite similar to the  
average distance $d_{j k}$ from $j$ to $k$, therefore the network's 
two perspectives (based on $i$ and $j$, respectively) will be quite similar.  
Consequently, $\Lambda(i,j)$ will 
be small if $i$ and $j$ belong to the same community and large if 
they belong to different communities.

\section{the algorithm} 
\label{sec:algorithm} 
 
We exploit the dissimilarity index to decipher the community 
structure of a network. After the distance matrix $\{d_{i j}\}$ and  
the dissimilarity indices for all the nearest-neighboring vertices 
$\{\Lambda(i,j)\}$ are obtained, the algorithm works as follows: 
 
\begin{description} 
 
\item[1.] Intially the whole network is just one single community.  
This community is 
assigned an upper dissimilarity threshold $\theta_{\rm upp}$ equalling 
to the maximum value of all the different dissimilarity indices. 
 
\item[2.] For each community, a resolution threshold parameter 
$\theta$ is introduced and is assigned the initial value $\theta_{\rm upp}$ 
of that community. The algorithm  is unable to discriminate between 
two nearest-neighboring vertices $i$ and $j$ when $\Lambda(i,j)\leq \theta$; 
if this happens, vertices $i$ and $j$ are marked as $``$friends".  
 
\item[3.] The $\theta$ value is decreased differentially. 
All edges in the community are examined to see whether two nearest-neighboring 
vertices are friends. Different friends sets of the community 
are then formed, each of which contains all the friends of the vertices in the 
 set. There may  
also be vertices in the community that do not have any friends.  
Each of these vertices is moved to  
the  friends set that has the strongest interaction with it.  
After this operation, vertices of the  community are distributed 
into a number of disjointed sets (this number may be unity).  
 
\item[4.] Each vertex in a subcluster should have 
stronger interaction with vertices within this subcluster than with vertices of any another 
subcluster of this community. To fulfill this requirement,  
we perform a local adjustment 
process: move each of the vertices that fail to meet this requirement 
to the friends set that has the strongest total interaction with it. This adjustment process 
is performed simultaneously for all these unstable vertices and is repeated until 
no unstable vertices remains.  
  
\item[5.] If vertices of the community remain together,  
the algorithm returns to step 3. If 
 they are divided into two or more sets, then the community under processing 
is assigned a lower dissimilarity threshold $\theta_{\rm low}$ equalling to 
the current $\theta$ value, and it is no longer considered. 
Each of the identified subsets of this community is regarded as a  
new (lower-level) community, with upper dissimilarity threshold $\theta_{\rm upp}$ 
equalling to the current $\theta$ value. 
The algorithm returns to step 2 to work with another identified community. 
 
\item[6.] After all the (sub)communities are processed, a dendrogram is drawn 
to demonstrate the relationship between different communities as well as the 
upper and lower dissimilarity thresholds of each community. The vertex set of  
each community is also reported.  
 
\end{description} 
 
The above procedure could be easily implemented with  
{\tt C++} programming language.  
The source code 
as well as the data for the examples studied in the following section  will be made 
publicly available \cite{zhouwebsite}.

\section{Applications} 
\label{sec:application} 
 
We test the performance of the above-mentioned algorithm by applying 
it first to two kinds of artificial networks and to  
three real-world networks. 
 
\subsection{Artificial random modular networks} 
\label{sec:artificial} 
 
To quantitatively compare with the work of Girvan and Newman \cite{girvan2002} 
the algorithm is first applied to a random modular network. 
The network has $128$ nodes, which are divided into 
$4$ modules of size $32$ each. Each vertex has 
on average $16$ edges connecting to other vertices, and on average  
$\overline{z}_{\rm out}$ of each vertex's edges  
are to vertices of other modules.  
All the edges are setup randomly with these two fixed expectation values. 
The present method is able to recover the modular structure 
of the network up to $\overline{z}_{\rm out}\simeq 7$. It  
slight outperforms the method of Girvan and Newman \cite{girvan2002}  
in performance. 
For example, working on an ensemble of random graphs with  
$\overline{z}_{\rm out}=6.0$ by the present method,  
on average only $4.5$ vertices are misclassified, each of which  
is assigned a cluster identity different from those of the majority of vertices of  
its module; 
while on average 
about $13$ vertices are misclassified by the method of  
Girvan and Newman \cite{girvan2002}.  

\begin{figure}  
\vspace*{1.0cm}
\hspace*{-6.0cm}\psfig{file=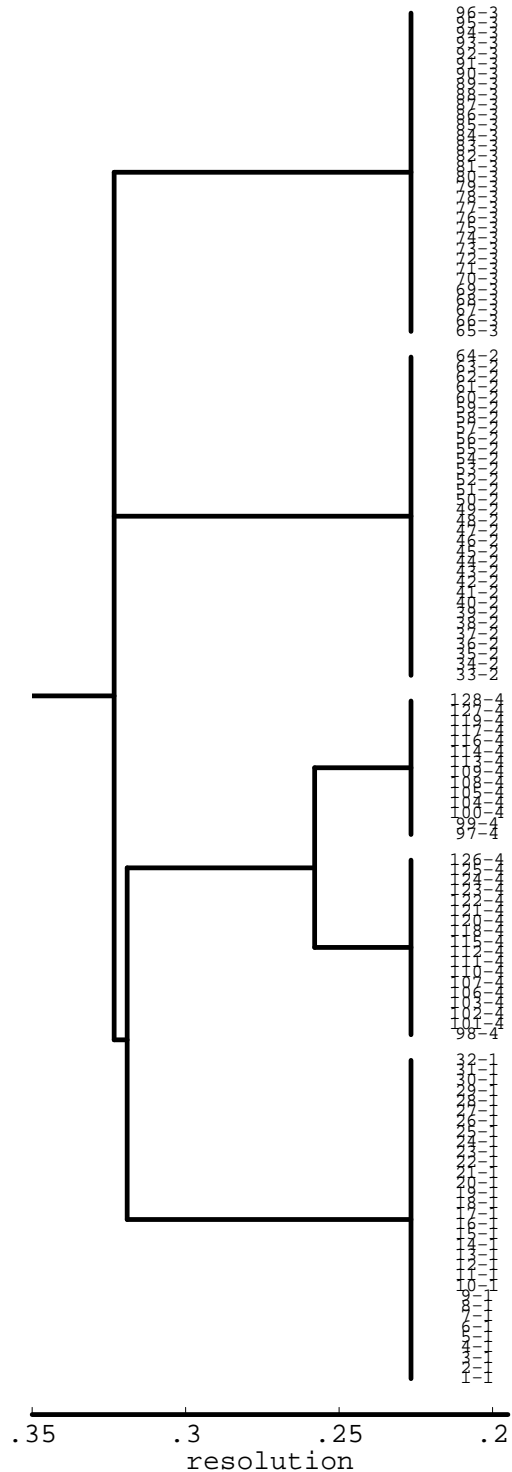,height=20.0cm} 
 
\caption{\label{fig:fig1} The community structure of a random 
modular network of $128$ vertices and $1067$ unweighted edges  
(see the main text for the rules how such a network is generated). 
Here and in following figures, in the pattern $xx$-$yy$, 
the number $yy$ after the hyphen denotes the group-identity 
of vertex $xx$ according to  information from other sources.} 
\end{figure} 
 
In figure \ref{fig:fig1}, the community structure of a randomly 
generated modular network with $\overline{z}_{\rm out}=6.0$ is 
demonstrated. When the resolution threshold is beyond $0.323$,  
the network as a whole could be regarded as 
a giant community. At resolution threshold $0.323$, however, $3$ subgroups 
suddenly emerge, with size $32$, $32$, and $64$, respectively. 
The first two communities correspond to two modules 
of the network, and the last one is the merge of the other two modules.  
At resolution threshold $0.319$, this later community  
again is divided into two subcommunities of $32$ vertices each, 
corresponding to the remaining two modules. 
At resolution threshold $0.258$, one of the modules of the 
network is found to fission into two subgroups of size $14$ and 
$18$, respectively. In this example, the designed four modules 
of the network correspond to the resolution range from  
$0.258$ to $0.319$. 
 
How to interpret the resolution parameters in the dengrograms 
such as that shown in figure \ref{fig:fig1}?  Take module $2$ and 
module $3$ as  examples. Figure \ref{fig:fig1} suggests that 
edges between these two modules have dissimilarity indices  
larger than $0.323$, while edges within these modules have 
dissimilarity indices $\simeq 0.227$. Therefore there is a 
large dissimilarity gap of about $0.1$ between an inter-modular edge 
and an intra-modular edge. 
 
It is noticeable that by the present algorithm, each community has 
certain range of stability. Subcommunities emerge only when the 
resolution threshold is lowered below certain level, and they  
emerge abruptly.  
 
\subsection{Regular hierarchy networks} 
\label{sec:hierarchy} 
 
\begin{figure*}
\hspace*{3.0cm}
\psfig{file=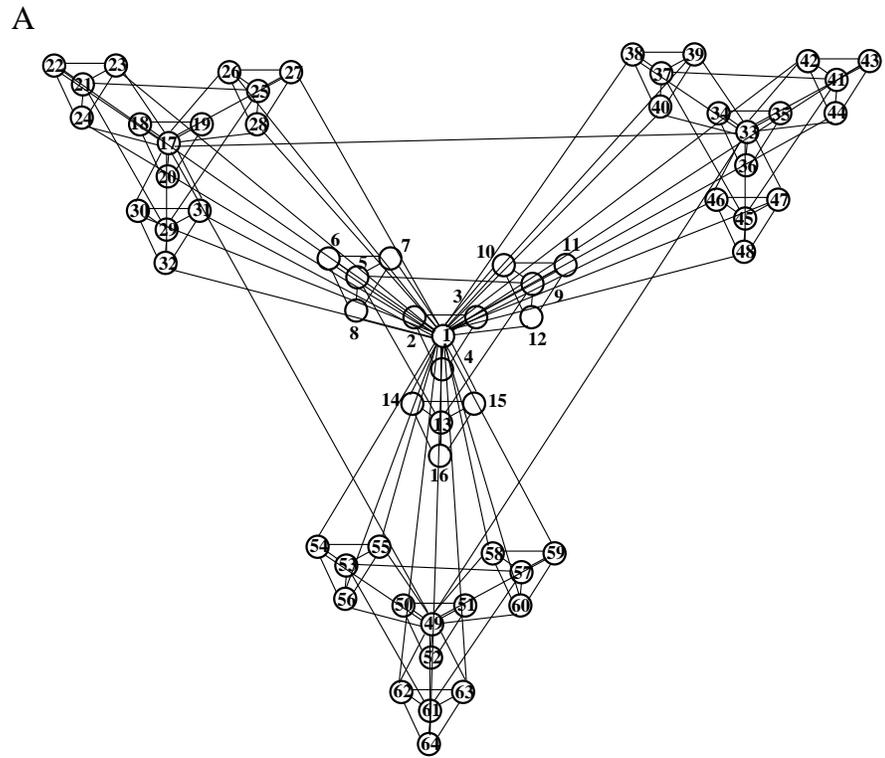,height=10.0cm} 

\vspace*{1.5cm}
\hspace*{-2.0cm}\psfig{file=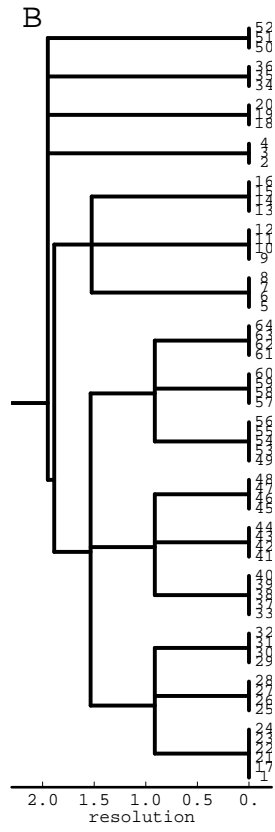,height=11.0cm} 
 
\caption{\label{fig:fig2} A hierarchy network \cite{ravasz2002}  
at level $n=2$ (A) and its community structure (B). }  
\end{figure*}

We analyze here the community structure of the model hierarchy 
network studied by Ravasz and coauthors \cite{ravasz2002}. The network 
is constructed by several steps \cite{ravasz2002}: At level $n=0$,  
a fully connected unit of four vertices is generated. At level $n=1$, three replicas of  
this unit are added and the external vertices of these replicas are 
connected to the central vertex of the $n=0$ unit, while 
the central vertices of the replicas are connected to each other. 
This replication-connection process could be continued to 
any desired level $n$. In figure \ref{fig:fig2}A such a 
network at level $n=2$ is shown. It was remarked \cite{ravasz2002} that conventional 
network clustering methods are unable to uncover the 
hierarchical structure of such a network. The present 
method, however, works very well:  
figure \ref{fig:fig2}B demonstrates the obtained community structure of the network 
figure \ref{fig:fig2}A.  
The hierarchy organization of the vertices in the network is largely reserved 
in \ref{fig:fig2}B. At resolution threshold 
$1.95$ the network is divided into $4$ subgroups of size 
$3$ and a giant component of size $62$.  
Later at resolution threshold $1.89$, this giant component again 
is fissioned into $2$ parts: one part has size $12$ 
and is further divided into $3$ subgroups of  
size $4$ at resolution  threshold $1.52$;  
the other part has size $50$, which, at 
resolution threshold $1.53$ further decomposes into  
$3$ subgroups of size $13$, $13$, and $14$, respectively. 
At resolution threshold $0.91$, each of these  three subgroups is further 
divided into $3$ subgroups.  

\subsection{The karate club network} 
\label{sec:zachary} 
 
\begin{figure} 
\vspace*{0.5cm}
\hspace*{-1.0cm}\psfig{file=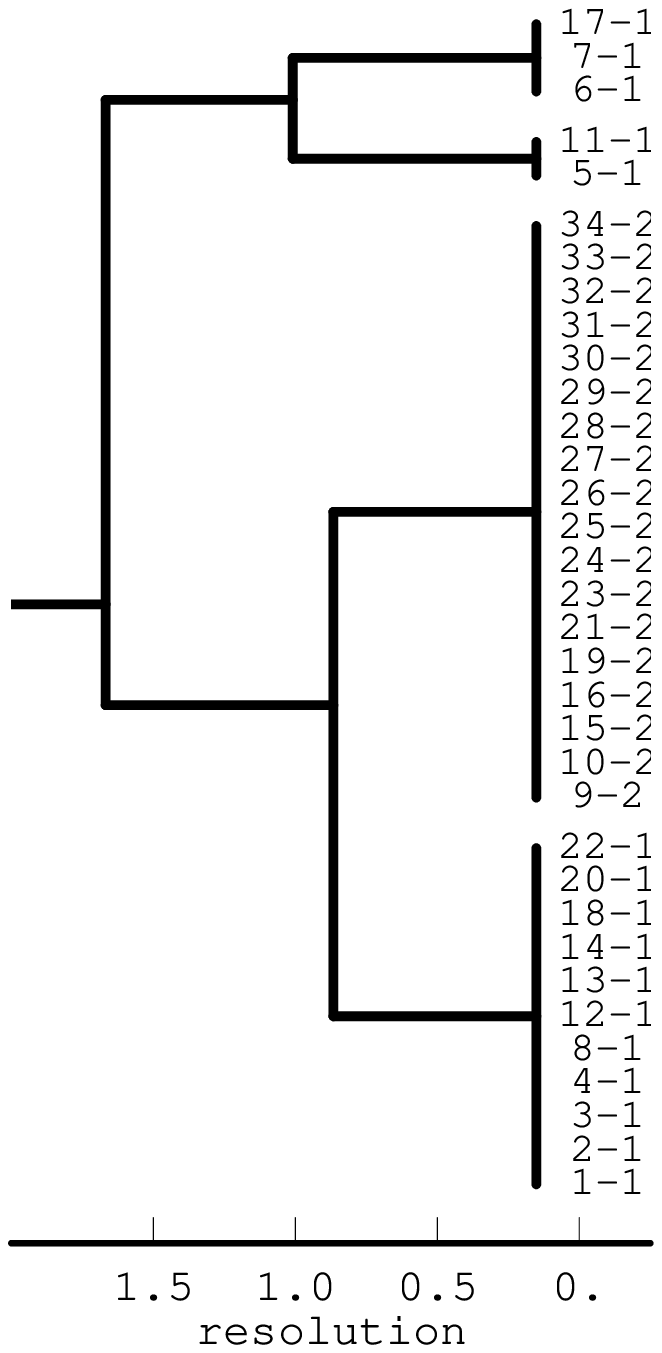,height=10.0cm} 
 
\caption{\label{fig:fig3} The community structure of 
the karate club network of Zachary \cite{zachary1977}.} 
\end{figure}

The karate club data \cite{zachary1977} examined in  
references \cite{girvan2002} and \cite{zhou2003} is re-evaluated 
here. This network is  weighted, each edge is assigned  a  
different strength. The present algorithm leads to the community structure 
of figure \ref{fig:fig3}. At resolution threshold $1.67$ the network 
decomposes into one small component of $5$ vertices and 
a large component of $29$ vertices.  At resolution threshold $0.87$,  
this large component further decomposes into two subgroups, 
One of which has $18$ members and the other has $11$ members.  
Comparison with the actual fission pattern is also shown in  
figure \ref{fig:fig3}. 

\subsection{The foot-ball team network} 
\label{sec:girvan1} 
  
\begin{figure} 
\hspace*{-1.5cm}\psfig{file=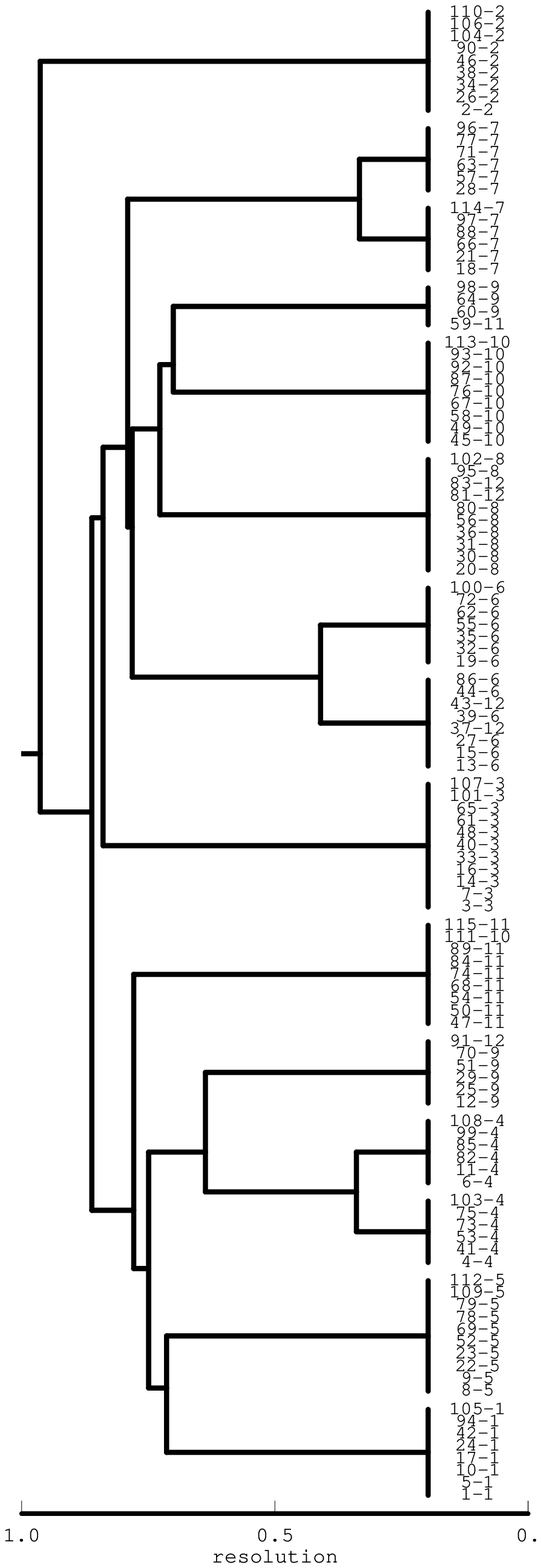,height=22.0cm} 
 
\caption{\label{fig:fig4} The community structure of 
the foot-ball network of Girvan and Newman \cite{girvan2002}.} 
\end{figure}

The foot-ball team network compiled by Girvan and Newman 
\cite{girvan2002} and studied in references 
\cite{girvan2002} and \cite{zhou2003} is re-investigated here. 
The present method results in the community structure 
of figure \ref{fig:fig4}. Each vertex's actual group-identity 
is also shown for comparison. 
In the resolution region between $0.41$ and $0.64$ there 
are $12$ communities according to the present algorithm. 
Of the $12$ actual groups, only members from group-$12$ 
are distributed  to other groups (with good reasons, 
because actually there are very few direct interactions  
between the five members of this cluster). Vertex $111$ 
are classified together with members of group-$11$, we have 
checked that this vertex has $8$ edges linking to group-$11$ 
and only $3$ edges to other groups. Vertex $59$ is classified together with  
members of group-$9$, we have also checked that it has 
stronger interaction with group-$9$ than with any another  
group.  

The organization of the different teams suggested by 
the present algorithm seems to 
be even better than the original organization.

\subsection{The scientific collaboration network} 
\label{sec:girvan2} 
 
\begin{figure} 
\vspace{0.5cm}
\hspace*{-2.5cm}\psfig{file=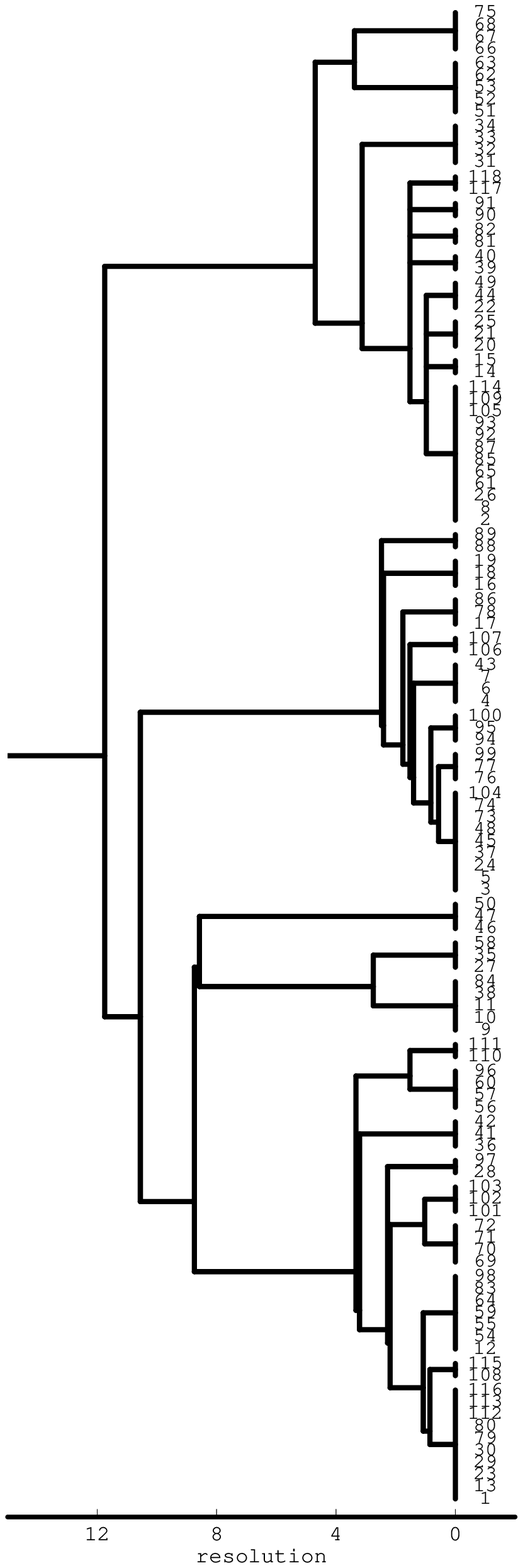,height=21.0cm} 
 
\caption{\label{fig:fig5} The community structure of 
the scientific collaboration network of Girvan and Newman 
\cite{girvan2002}.} 
\end{figure}

The scientific collaboration network compiled by Girvan and 
Newman \cite{girvan2002} and examined in references \cite{girvan2002} and  
\cite{zhou2003} is also re-examined. This network is 
also weighted. The present method suggests 
a community structure shown in figure \ref{fig:fig5}. 
In accordance with the actual situation, on the global scale,  
the network clearly has $3$ giant  communities of comparable 
sizes.  
Each of these giant communities  could further be decomposed into  
several subcommunities when the resolving power is increased. 

\subsection{The protein interaction network of yeast} 
\label{sec:yeast} 
 
The protein interaction network of yeast is constructed based on 
the data reported in references \cite{xenarios2000} and 
\cite{deane2002}, it contains $1471$ proteins and $2770$ edges 
(protein-protein physical interactions). This network has already 
been studied in reference \cite{zhou2003}; here we constructed a 
reduced interaction network based on the original one. First, 
self-connection is removed; second, proteins which are connected to 
the network by only  
one edge are removed. The second step is continued until no proteins 
of degree one remains. The reason to remove all the proteins of degree one  
is that, according 
to the idea of Girvan and Newman \cite{girvan2002}, a vertex that 
is connected to the network by just one edge should be in the same 
community as its nearest-neighboring vertex,   therefore its 
status need not to be considered separately. Of cause, we have 
checked that actually identical results are obtained when the 
network-reduction process is not performed. The reduced network 
contains $871$ proteins and $2043$ unweighted interactions (edges). 

\begin{figure*} 
\hspace*{1.0cm}\psfig{file=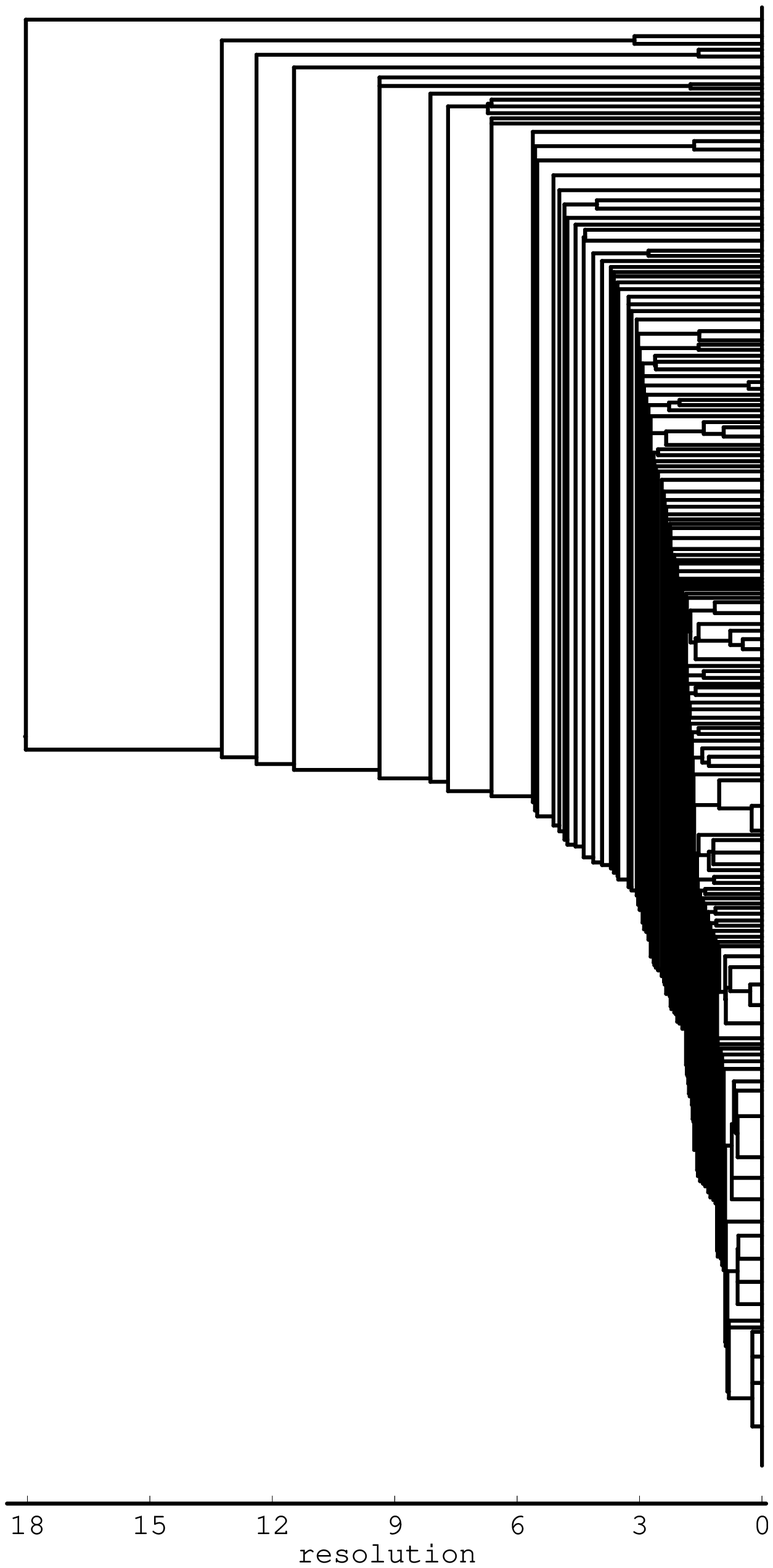,height=22.0cm} 
 
\caption{\label{fig:fig6} The community structure of  
the reduced protein-protein interaction network of yeast.} 
\end{figure*}

The community structure of this network is demonstrated in figure  
\ref{fig:fig6}. It seems to 
be strikingly different from those of  the other networks studied in 
this paper. At the resolution range between $\sim 1.5$ to $18.0$ many 
small communities appear, but the network is dominated by just one  
large cluster of size proportional to the total size of the network. 
This is in accordance with reference \cite{zhou2003} where the original 
network was decomposed into one large component and several small 
components.  
When the resolution threshold  is decreased below  $1.5$, 
the largest cluster is 
divided into several subclusters of comparable sizes. The biological 
significance of such a community structure is yet to be 
investigated. 
 
Based on the community structure shown in figure \ref{fig:fig6}, we 
can construct clusters of proteins that might be of  
biological significance. Here we just show three examples of such  
protein clusters, corresponding respectively to higher, medial, and lower 
resolution thresholds.

\begin{figure} 
\psfig{file=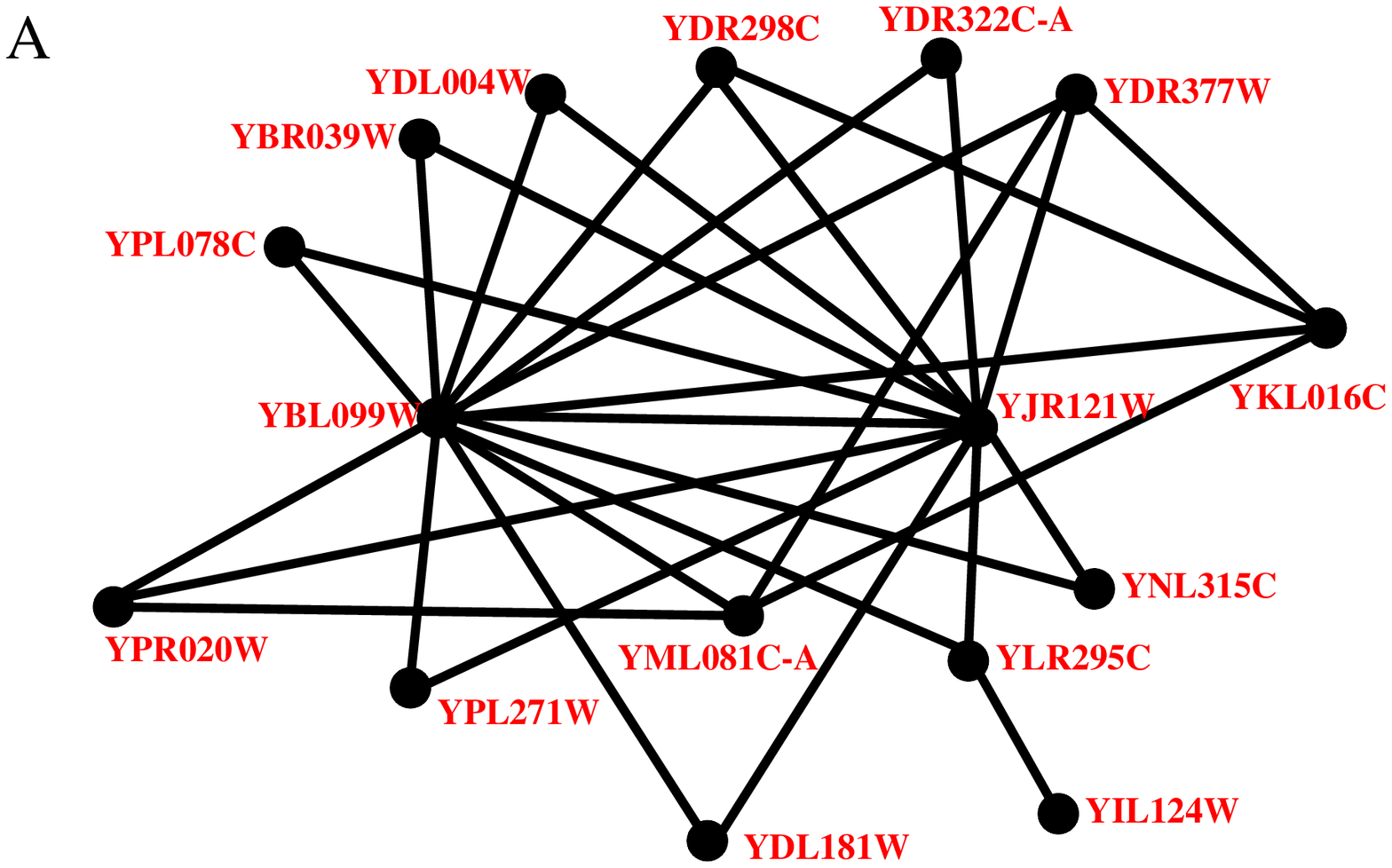,height=5.5cm} 

\vskip 1.0cm
\psfig{file=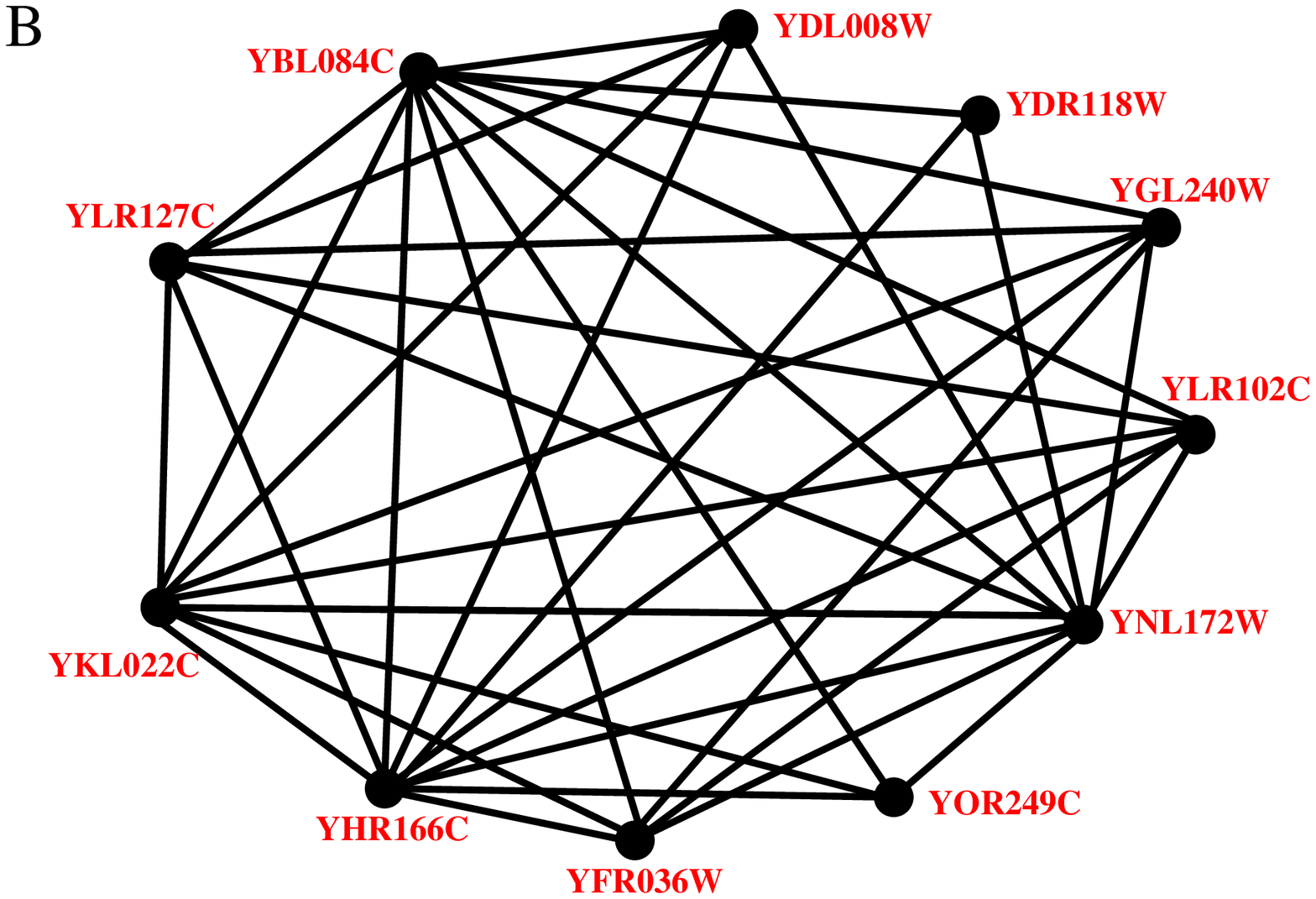,height=5.5cm} 

\vskip 1.0cm
\psfig{file=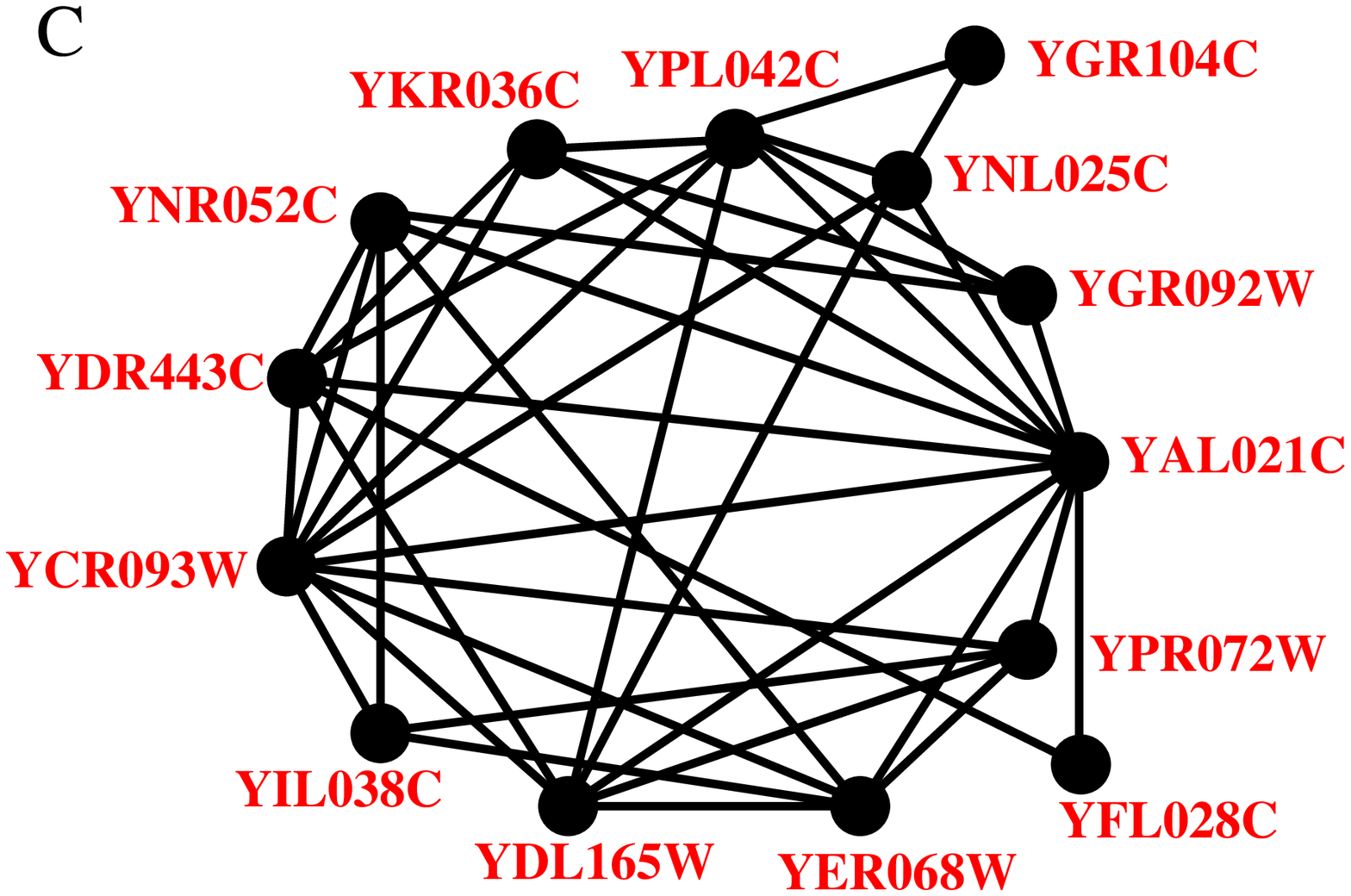,height=5.5cm} 
 
\caption{\label{fig:fig7} Examples of proteins clusters 
identified according to the community structure of figure 
\ref{fig:fig6}.} 
\end{figure}

The first example is a cluster which appears at resolution threshold 
$18.04$. It contains $16$ proteins and $33$ edges, and has the structure 
shown in figure \ref{fig:fig7}A. This cluster is stable, namely 
that each vertex in it is more connected to vertices in this cluster than 
to vertices outside; and it has no further subcommunity structure. 
According to the protein interaction databank \cite{xenarios2000,bairoch2000}, $15$ of 
these proteins are all involved in ATP synthesis process in yeast. 
They may form a very important part of yeast's mitochondrial ATPase 
complex. One protein of this cluster, {\tt  YIL124W}, is a hypothetical 
membrane protein. Because this last protein has only one interaction 
with other members of the cluster, it may not have similar biological  
functions as the other members.

The second example is a  
cluster which appears at resolution threshold $5.11$. It 
contains $11$ proteins and $38$ edges, and has the structure 
shown in figure \ref{fig:fig7}B. This cluster is also stable and has no 
further structure. According to the protein interaction databank 
\cite{xenarios2000,bairoch2000}, 
among these $11$ proteins, 
{\tt YBL084C, YFR036W, YHR166C, YKL022C} are known to be cell division 
control proteins;  
{\tt YGL240W} plays a role in cell cycle and mitosis;  
{\tt YDR118W, YNL172W, YOR249C} probably are membrane proteins; and 
{\tt YLR127C, YDL008W, YLR102C} 
are hypothetical proteins whose functions remain 
to be determined. It is quite likely that all the proteins in this cluster are 
closely involved in  
cell division and membrane fission process. We anticipate that 
the  three hypothetical proteins  of this cluster will also have similar 
biological functions.  
 
The third example is a cluster which appears only when the resolution  
threshold is 
refined to below $0.88$. It contains $14$ proteins and $41$ protein-protein 
interactions. This cluster is also stable and has no 
further structure. The interaction pattern of this cluster is demonstrated 
in figure \ref{fig:fig7}C. Among these $14$ proteins, according to the 
protein interaction databank \cite{xenarios2000,bairoch2000},  
{\tt YCR093W, YPR072W, 
YDL165W, YER068W, YIL038C} are general negative regulator of 
transcription subunits;  
{\tt YAL021C} is a glucose-repressible alcohol dehydrogenase 
transcriptional effector; {\tt YNR052C} is a ubiquitous transcription 
factor; {\tt YDR443C, YGR104C} are suppressors of RNA polymerases; 
{\tt YNL025C} is the RNA polymerase II holoenzyme cyclin-like subunit;  
{\tt YPL042C} is the meiotic mRNA stability protein kinase {\tt UME5}; 
{\tt YGR092W} is the cell cycle protein kinase {\tt DBF2};  
and {\tt YKR036C} and 
{\tt YFL028C} are two hypothetical proteins.  
It is quite likely that this cluster is mainly involved in RNA transcription 
process and we also anticipate that the two hypothetical proteins of this cluster 
are strongly related with this biological function.

To conclude this subsection, we stress that,  
based on the community structure 
of figure \ref{fig:fig6} many clusters of proteins can be constructed. 
Here we have mentioned just three examples. 
These identified protein clusters could help 
researchers to assign possible biological functions to hypothetical  
proteins, and could also suggest possible proteins that may be involved 
in carrying out a particular biological reaction. 
 
\section{conclusion and discussion} 
\label{sec:conclusion}

In our earlier work \cite{zhou2003}, the distance between two vertices of  
a graph is defined as the average number of steps a Brownian particle takes 
to move from one vertex to the other. Based on 
this distance measure, in the present work we  
define a dissimilarity index to signify to what extent two nearest-neighboring 
vertices will be different from each other. We observe that vertices 
belonging to the same group usually have very small dissimilarity indices  
between them, while 
vertices of different communities usually have large dissimilarity indices 
between them.  
The observation leads naturally to an algorithm of network clustering. 
We applied this method to several artificial networks and also to different  
real networks in social and biological systems and satisfactory results 
are obtained. Different clusters of a network obtained by our method are 
characterized by a range of resolution threshold.

The examples studied by us in this paper  
suggest that our algorithm is 
very promising in identifying the community structure of a complex networked 
system. Why it works? Maybe it is because of the following reasons. 
First, the vertex-vertex distance measure has taken into account the 
topological structure of the network as well as the local connections of the 
network. The distances from one vertex to all the other vertices of the 
network actually give a perspective of the whole network viewed from this 
vertex.  Second, the dissimilarity index defined by equation (\ref{dsi}) 
compares the perspectives  viewed from two nearest-neighboring vertices.  
It is intuitively appealing to assume that the perspectives of the different 
vertices of the same community are similar to each other while those of 
vertices of different communities will be quite different.

It is anticipated that the present work will find applications in the  
field of complex networks, as well as in  the fields of sociological  
and biological sciences.

 \section*{Acknowledgement} 
 
This research is made possible by a  
post-doctoral fellowship from the Max-Planck 
Society. The author is grateful to Professor Reinhard Lipowsky 
for his constant support.  
 

\end{document}